\documentclass[a4wide,12pt]{article}
\usepackage[margin=1.2in]{geometry}
\usepackage{amsmath,amssymb,latexsym}
\usepackage{svg}

\usepackage{mathptmx} %% !
\usepackage{mathtools}
\usepackage[style=numeric-comp,sorting=none, backend=bibtex]{biblatex}
\bibliography{refs}

\def \sec{\begin{section}}
\def \esec{\end{section}}

\newtheorem{proposition}{Proposition}
\newtheorem{definition}{Definition}

\renewcommand{\tilde}{\widetilde}
\renewcommand{\bar}{\overline}

\def \Oc {\mathcal{O}}

\def \Sc {\mathcal{S}}

\def \ra {\rightarrow}

\def \beq { \begin{equation}}
\def \eeq {\end{equation}}
\def \swap {\text{SWAP}}

\def \OTOC {\text{OTOC}}

\DeclareMathOperator*{\Tr}{Tr}
\def \id {\mathbb{I}}

\def \l {\left}
\def \r {\right}

\def \bra {\langle}
\def \ket {\rangle}

\usepackage{authblk}

\def \SWAP {\text{SWAP}}

\usepackage{comment}

\begin{document}
\title{\Large How to distinguish chaos from integrability using OTOC}
\author[1,2]{Alexey Milekhin$^*$}
\author[3,4]{Anders Wallberg\footnote{Equal contribution.}}
\affil[1]{\normalsize Institute for Quantum Information and Matter, California Institute of Technology, Pasadena, CA 91125, USA}
\affil[2]{\normalsize Department of Physics and Astronomy, University of Kentucky, Lexington, KY 40506, USA \footnote{milekhin@uky.edu}}
\affil[3]{\normalsize Department of Theoretical Physics, CERN, 1211 Meyrin, Switzerland}
\affil[4]{\normalsize Laboratory for Theoretical Fundamental Physics, EPFL, 1015 Lausanne, Switzerland}
\date{}

\maketitle

\begin{abstract}
    The decay of out-of-time-ordered correlators (OTOCs) to a small residual value signals the scrambling of quantum information. This behavior is generic: it occurs in both chaotic and interacting integrable many-body systems, and therefore cannot by itself distinguish between them. In this paper we analytically study the late-time saturation value of a closely related observable, operator purity, and show that its scaling with system size robustly distinguishes chaotic from integrable systems. Our approach is based on the Mazur bound, which provides concrete scaling estimates and makes explicit why the locality of integrable charges matters. We support our findings with numerical simulations of one-dimensional spin chains.
\end{abstract}

\section{Introduction}
\label{sec:intro}
Unlike the well-developed theory of classical chaos, quantum chaos remains elusive. The general expectation is that integrable theories possess many additional conservation laws, whereas chaotic theories do not. In the quantum setting, this statement comes with considerable fine print: in any quantum system, the projectors onto energy eigenstates commute with the Hamiltonian, but they are highly non-local. The central question is therefore how to detect genuinely local or quasi-local conserved charges operationally, using physical observables. One of the earliest diagnostics is level repulsion \cite{wigner,dyson,bgs} in the spectra of chaotic systems. However, it requires observing the system for times exponentially long in the system size and obscures the role of charge locality.

\begin{figure}[h!]
    \centering
    \includegraphics{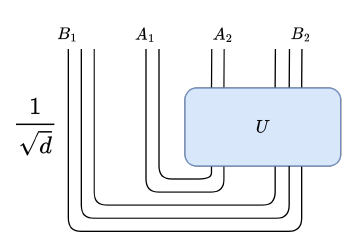}
    \caption{The Choi--Jamiolkowski state $| U \ket $ and its partition into a subsystem $A$ and its complement $B$. There are two copies of each: $A_1, B_1$ at time zero and $A_2, B_2$ after the evolution $U$. This is why $\Tr \rho_{A_1 A_2}^2$ is, in some sense, entanglement "in time" \cite{Brukner:2004egd,Hosur:2015ylk,Cotler:2017anu,Doi:2023zaf,Milekhin:2025ycm}.}
    \label{fig:CJ}
\end{figure}

Over the past decade, out-of-time-ordered correlators (OTOCs) have attracted attention across many areas of physics, including quantum chaos (we refer to \cite{Garcia-Mata:2022voo} for a review). They take the form (the subscript indicates the subsystem supporting the operator)
\beq
\OTOC(t,O_A, O_B)_{\rho} \equiv \Tr (\rho O_A(t) O_B(0) O_A(t) O_B(0) )\,.
\eeq
Alternatively, one can average this over all operators supported in a given subsystem \cite{Hosur:2015ylk}. This yields a quantity, the so-called operator purity (or more generally operator R\'enyi entropy), that depends only on the evolution operator and the choice of subsystem. At infinite temperature it can be written as
\beq
\Tr \rho_{A_1, A_2}(t)^2 = \frac{1}{d d_A d_B} \sum_{O_{A_1}, O_{B_1}} \OTOC(t,O_{A_1}, O_{B_1})_{\id/d}\,,
\eeq
where the sum runs over all Pauli strings supported in the regions $A_1$ and $B_1$, and $\rho_{A_1 A_2}$ is the reduced density matrix of the Choi--Jamiolkowski state associated with the evolution operator $U$ (Figure \ref{fig:CJ}). Throughout the paper, $B=\bar{A}$ denotes the complement of $A$, $d_A = 2^{L_A}$ is the dimension of subsystem $A$, and $d$ is the dimension of the full Hilbert space.

OTOCs probe properties of quantum systems that go beyond conventional transport and hydrodynamics. One of their original motivations \cite{Larkin:1969abc} was to define a quantum Lyapunov exponent from the early-time behavior of the OTOC, measuring the sensitivity of a system to initial conditions. This can be done quantitatively, however, only in systems with a large semiclassical parameter, such as large-$N$ holographic systems \cite{Shenker:2013pqa,Maldacena:2015waa}.
More generally, OTOCs probe the scrambling (delocalization) of information \cite{Xu:2019lhc}, signaled by the OTOC starting from a large value at $t=0$ and decaying to a small value at late times. This behavior occurs in most many-body quantum systems \cite{Iyoda:2017pxe}, as long as they are interacting (OTOCs do not decay, for example, for free fermions \cite{Lin_2018}).

In this paper, we revisit the question of whether operator purity can distinguish chaos from integrability. We argue that it can, through its \emph{late-time saturation} value rather than through the early-time Lyapunov regime. Since the operator entropy is a well-studied quantity that arises in contexts ranging from quantum advantage \cite{google2025observation} to the black hole information paradox \cite{Hartman:2013qma}, we hope that our results will be useful beyond the question of chaos versus integrability.

Our approach to the residual value immediately explains why the locality or quasi-locality of the conserved charges is important. We use the fact that $\Tr \rho_{A_1 A_2}^2$ is simply the two-point function of the operators $\SWAP_A$ in the \textit{doubled system}, which exchange the two copies of $A$:
\beq
\Tr \rho_{A_1 A_2}^2 = \frac{1}{d^2} \Tr(\SWAP_A(0) \SWAP_A(t)) .
\eeq

Mathematically, we study the infinite-time average
\beq
\label{eq:time_avg}
\overline{\Tr \rho_{A_1 A_2}^2} \equiv  \lim_{T \ra +\infty} \frac{1}{2T} \int_{-T}^{+T} dt \ \Tr \rho_{A_1 A_2}^2(t).
\eeq

Physically, however, one expects OTOCs to saturate at times much shorter than the Heisenberg time (the inverse level spacing), which would make the distinction between chaos and integrability observable in practice. Specifically, saturation should occur \footnote{We refer to \cite{Dymarsky:2018ccu} for the discussion on possible long-time hydrodynamic tails. However, they still decay in a time polynomial in the system size.} at the Thouless time, which for diffusive systems is $L^2/D$, with $L$ the linear size and $D$ the diffusion constant \cite{DAlessio:2015qtq}; we provide some limited numerical evidence for this.

Mazur \cite{mazur1969non} derived a general lower bound on such infinite-time averages of two-point functions in terms of \textit{any} set of conserved charges $Q_k$. For SWAP operators and Hilbert--Schmidt orthogonal traceless charges $Q_k$, it gives
\beq
\overline{\Tr \rho_{A_1 A_2}^2}  \ge  \frac{1}{d_A^2} + \frac{1}{d^2} \sum_k \l( \frac{ \Tr_A[\Tr_{\bar{A}}(Q_k) \Tr_{\bar{A}}(Q_k) ]}{(\Tr Q_k^2)} \r)^2 \,.
\label{eq:intro_mazur}
\eeq
We derive this inequality in the main text. The first term is the answer for a Haar-random unitary,\footnote{Up to an exponentially small correction $d_A^2/d^2$; throughout the paper we assume $L_A < L/2$. We comment on this further in Sec.~\ref{sec:uni}.} while the remaining terms come from the projection onto the span of the conserved charges $Q_k$. This formula makes quantitative the intuition that conserved charges obstruct scrambling.
It also makes manifest why the locality of the $Q_k$ matters. Projectors onto energy eigenstates $| E_k \ket \bra E_k |$ give an exponentially small contribution, because the expression in parentheses reduces to the purity of subsystem $A$, which for excited states is typically exponentially small in the size $L_A$ of $A$. In contrast, as we will see, local integrable charges contribute at order $(L_A/L)^2$.

This leads directly to our main results. Let us assume one spatial dimension and charges $Q_k$ built from Pauli strings; our results generalize easily to other cases. In the presence of an extensive number of quasi-local conserved charges (defined below), the quantity $d_A^2 \overline{\Tr \rho_{A_1 A_2}^2}$ grows \textit{at least linearly} with the size $L_A=|A|$ at fixed $L_A/L$:
\beq
\label{eq:int_val}
\boxed{\text{integrable:} \quad d_A^2 \times \overline{\Tr \rho_{A_1 A_2}^2} \ge L_A \times I(L_A/L) }
\eeq
In the main text, we explain how to determine the function $I(L_A/L)$, which depends on the microscopic details of the theory. The reason for this linear growth is simple: the partial trace $\Tr_{\bar{A}} Q_k$ selects only those charge densities that fit entirely inside the interval $A$, since all others have vanishing partial trace. In many integrable systems, for example the Heisenberg XYZ chain, there are of order $L_A$ such charges, because $Q_k$ is a sum of Pauli strings acting on $k$ adjacent qubits.

In contrast, for a non-integrable system we expect the Hamiltonian $H$ to be the only conserved charge, so that this quantity depends only on $L_A/L$:
\beq
\label{eq:chaos_val}
\boxed{\text{chaotic:} \quad d_A^2 \times \overline{\Tr \rho_{A_1 A_2}^2} \ge 1 + \frac{d_A^2}{d^2} \l( \frac{ \Tr_A[\Tr_{\bar{A}}(H) \Tr_{\bar{A}}(H) ]}{(\Tr H^2)} \r)^2 = C \l( \frac{L_A}{L} \r) }
\eeq
In Section \ref{sec:energy} we discuss when this bound, supplemented by higher powers of the Hamiltonian, is actually saturated in chaotic systems.

\iffalse
\begin{table}[h!]
    \centering
    \begin{tabular}{c|c}
    system & $d_A^2 \times \bar{ \Tr \rho_{A_1 A_2}^2}$ \\
    \hline
      2-design   &  $1 + \Oc(d_A^2/d_B^2)$ \\
      Only energy conservation   & $\l( 1+  \cfrac{L_A^{{\color{red} 2}}}{L^2} + \Oc((L_A/L)^3) \r) + \Oc(d_A^2/d_B^2) $\\
      Integrable hierarchy & $\l( 1+ \cfrac{L_A^{{\color{red} 3}}}{3L^2} + L_A \times \Oc((L_A/L)^3) \r) + \Oc(d_A^2/d_B^2) $
    \end{tabular}
    \caption{Lower bound for the second Renyi entropy $\Tr \rho_{A_1 A_2}^2$. Notice an extra power of $L_A$ for the integrable hierarchy. }
    \label{tab:bound}
\end{table}
\fi

\paragraph{Relation to prior works}
Our paper was inspired by \cite{Huang:2017fng} studying a similar bound on individual OTOC coming from energy conservation and observing large deviations from Haar-random values.
Ref. \cite{Kudler-Flam:2021kfl} investigated the impact of conserved charges on scrambling using operator entanglement. However, because of a different region arrangement, they did not observe a separation between integrable and chaotic cases.
Results in our paper provide an analytical explanation for the numerical scaling results observed for the long-time average \cite{Styliaris:2020tde,Anand:2021xet,Barch:2023btu} of $\Tr \rho_{A_1 A_2}^2$ and its fluctuations \cite{Fortes:2019frf}.

\paragraph{Notation}
\begin{itemize}
    \item Latin letters from the beginning of the alphabet ($A,B,C,\dots$) denote subsystems.
    \item $|A| \equiv L_A$ denotes the size of a subsystem (e.g.\ its number of qubits).
    \item $d_A$ is the dimension of the corresponding Hilbert space ($d_A = 2^{|A|}$ for qubits).
    \item $L$ and $d$ denote the size and dimension of the whole system.
    \item Latin letters from the middle and end of the alphabet ($I,Y,O,\dots$) denote operators.
    \item $H$ is the Hamiltonian, and $Q_k$ denote other conserved quantities.
\end{itemize}

\section{Mazur bound}
Consider an autocorrelation function of the form
\beq
\bra Y(t) Y(0) \ket = \Tr(\rho Y(t) Y(0)).
\eeq
If $\rho$ is diagonal in the energy basis, it is elementary to show that the time average is non-negative for \textit{any} Hermitian operator $Y$:
\beq
\bar{\bra Y(t) Y(0) \ket} \equiv \lim_{T \ra +\infty} \frac{1}{2T} \int_{-T}^{+T} dt \bra Y(t) Y(0) \ket \ge 0\,.
\eeq
Mazur \cite{mazur1969non} obtained the following bound, which we review in the Supplementary Material, Sec.~\ref{app:mazur}:
\begin{equation}
\label{eq:thebound}
    \bar{\bra Y(t) Y(0) \ket} \ge \sum_{ij} \bra Y Q_i \ket M_{ij} \bra Q_j Y \ket\equiv \text{B}(\left\{Q_k\right\}_{k=1}^N)\,,
\end{equation}
where $M_{ij}$ is the inverse of the matrix $\bra Q_i Q_j \ket$.

Throughout the paper, $B(\{Q_k\})$ denotes the bound generated by a set of operators $Q_k$. It would be very interesting to study non-commuting operators $Q_k$ and/or a non-trivial state $\rho$. In this paper, however, we restrict to commuting operators $Q_k$ and the maximally mixed (infinite-temperature) state, so that $\bra \cdot \ket$ is the normalized trace $\frac{1}{d} \Tr( \cdot )$.
Similarly, the normalized partial trace over a subsystem $A$ is denoted by $\bra \cdot \ket_A = \frac{1}{d_A} \Tr_A(\cdot)$.

In integrable systems, the conserved charges tend to be non-local, so a generic local observable $Y$ overlaps with only a small fraction of them. Can one construct operators that overlap with a finite fraction of the charges? For a single copy of the system, this does not seem possible, which is why we consider two copies and study SWAP operators. The overlap with a SWAP is generically non-trivial, because
\beq
\label{eq:swap_overlap}
\bra \swap_A X \otimes Y \ket = \frac{1}{d^2} \Tr_A \l[ \Tr_B(X) \Tr_B(Y) \r]\,,
\eeq
and for $X=Y$ this is proportional to $ \Tr_A \l[ \Tr_B(X) \Tr_B(X) \r]$, which is non-negative.

An immediate observation is that the residual value of the operator entanglement is sensitive not only to the conserved charges of the Hamiltonian $H$, but to the \textit{conserved charges of the doubled system} with Hamiltonian $H \otimes \id + \id \otimes H$. What are these charges?
Clearly, the evolution commutes with $Q_k \otimes Q_l$. In addition, there is the identity $\id$ and the total swap $\swap$ between the copies, as well as the products $(\swap)(Q_k \otimes Q_l)$. The swaps ensure that the bound is symmetric under $ A \leftrightarrow B$.
One can also use the powers $Q_k^2, Q_k^3, \dots$ and other symmetries, such as reflections and translations, some of which are included in the Supplementary Material. In the main text, we concentrate on simple operators, and defer a more detailed study of conserved charges in replicated systems to a future paper \cite{wip}.

Using the orthogonal set of charges $\{ \id \otimes \id, Q_k \otimes Q_k \}$ in the Mazur bound (\ref{eq:thebound}) together with the overlap formula (\ref{eq:swap_overlap}), we will immediately arrive at the bound (\ref{eq:intro_mazur}) quoted in the Introduction.
The only slightly surprising aspect of this computation is the inclusion of the identity $\id \otimes \id$, which is usually not regarded as a conserved charge. It nevertheless produces the important $1/d_A^2$ contribution to the bound, which we discuss in more detail in the next Section.

\section{Bounds on operator R\'enyi entropy}
Before making a general statement in Section \ref{sec:formal}, we consider a few examples.
\subsection{Universal bounds}
\label{sec:uni}
\paragraph{Identity}
Every quantum system has at least one conserved charge, the identity $\mathbb{I}$. For any operator $X$, the bound obtained from the identity is
\begin{equation}
    \overline{\bra X(t)X(0)\ket}\geq\bra X(0)\ket^2\,,
\end{equation}
which, in a time-translation-invariant state, is the statement that the time average of the \textit{connected} two-point function is non-negative. For the SWAP operator,
\begin{equation}
    \bra \swap_A \ket=\frac{d_B^2d_A}{d^2}=\frac{1}{d_A}\,,
\end{equation}
so the Mazur bound from the identity, $\text{B}(\mathbb{I})$, is
\begin{equation}
    \overline{\bra\swap_A(t)\swap_A(0)\ket}\geq \text{B}(\mathbb{I})=\frac{1}{d_A^2}\,.\label{idbound}
\end{equation}
This bound is expected, since $\rho_{A_1 A_2}$ is a density matrix on a Hilbert space of dimension $d_A^2$.

\paragraph{SWAP}
By construction, the swap operator of the entire system is also conserved. One finds
\begin{equation}
    \bra \swap_A\swap\ket=\bra\swap_B\ket=\frac{1}{d_B}\,,\qquad \bra\swap\,\swap\ket=1\,,
\end{equation}
so the purely diagonal contribution of the swap is
\begin{equation}
    \text{B}(\swap)=\frac{1}{d_B^2}\,.\label{swapbound}
\end{equation}
Combining it with the contribution of the identity, and using $\left<\swap\right>=\frac{1}{d}$, we obtain
\begin{equation}
    \text{B}(\{\mathbb{I},\text{SWAP}\})=\frac{2-d_A^2-d_B^2}{1-d_A^2d_B^2}\sim\frac{1}{d_A^2}+O(d_B^{-2})\,.
\end{equation}
This is the exact operator purity for Haar-random evolution, which preserves only the identity and the SWAP.

\paragraph{Translations} For translation invariant systems, we also analyze the contribution of the translation operator in the Supplementary Material, Section \ref{app:trans}, and find that the correction scales at most as $1/d_A^2$.

\paragraph{The size of $A$}
Before discussing the bounds from energy and other conserved charges, let us comment on the size of the region $A$.
Formally, all the bounds below remain valid when $A$ is larger than half of the system, but they then become very weak: the contribution of the total SWAP shows that $d_A^2 \overline{\Tr \rho_{A_1 A_2}^2}$ exceeds $d_A^2/d_B^2$. This contribution is exponentially small for $L_A<L/2$, but exponentially large for $L_A > L/2$. We therefore work formally with $L_A < L/2$, which allows us to neglect the contribution arising from SWAP.

\subsection{Energy Bounds}
\label{sec:energy}
We now turn to less universal bounds, starting with energy conservation. We assume that the Hamiltonian is translation invariant and has the form
\begin{equation}
    H=\sum_x h_x\,,\label{eq:Hamil}
\end{equation}
where $h_x=P^x\,h_0\,P^{-x}$, $P$ is the translation operator, and $h_0$ has a finite interaction length $\ell$ (it acts on $\ell$ consecutive sites) and satisfies
\begin{equation}
    \Tr(h_0)_{x\in \ell}=0\,,
\end{equation}
\textit{i.e.} all partial traces of $h_0$ vanish.

The contribution of this operator to the Mazur bound is easy to compute:
\beq
B(H \otimes H) = \frac{1}{d^2} \frac{\l( \Tr_A \l[ \Tr_B H \Tr_B H \r] \r)^2}{\l( \Tr H^2 \r )^2 } = \begin{cases}
\cfrac{1}{d_A^2} \cfrac{(L_A - \ell+1)^2}{(L)^2}, \quad \ell \le L_A \\
0, \quad \ell > L_A.\label{eq:BHH}
\end{cases}
\eeq
The factors $L_A-\ell+1$ arise because only the diagonal terms $h_x^2$ contribute. Since $\ell$ is typically fixed, for large $L_A$ and $L$ this is simply a function of $L_A/L$.

\paragraph{Saturation}
When can we expect the Mazur bound to be saturated?
Suzuki \cite{suzuki1971ergodicity} proved that, for a non-degenerate spectrum, the Mazur bound obtained from the powers of the Hamiltonian $\{ \id, H, H^2, H^3, \dots \}$ is an equality. The reason is that for a non-degenerate spectrum the time average projects onto $\text{span}( | E_1 \ket \bra E_1|, | E_2 \ket \bra E_2|, \dots)$, which coincides with $\text{span}(\id, H, H^2, \dots)$.

In our case, the relevant Hamiltonian is that of the doubled system, $H \otimes \id + \id \otimes H$. It is always degenerate because $|E_n \ket \otimes | E_m \ket$ has the same energy as $|E_m \ket \otimes | E_n \ket$. So we will need a different condition, usually called the ``non-resonance'' condition in the literature:
\beq
E_k + E_l \neq E_m + E_n \quad \text{unless} \quad E_k = E_m, E_l=E_n \ \text{or} \ E_k = E_n, E_l=E_m .
\eeq
This condition appears in many proofs in the many-body literature. The time average then projects onto \cite{Anand:2021xet} $\text{span}(|E_n \ket \bra E_n|  \otimes | E_m \ket \bra E_m| \cup |E_n \ket \bra E_m|  \otimes | E_m \ket \bra E_n| )$, which is the span of $H^a \otimes H^b$ and $(H^a \otimes H^b) \swap$ over all powers $a,b$. In this case we expect our bound to become an equality.
In the Supplementary Material, Sec.~\ref{app:Hpower}, we analyze the powers of the Hamiltonian and find that they contribute higher powers of $L_A/L$.
We therefore expect the operator purity of chaotic systems to only be a function of $L_A/L$, which our numerical results in Section \ref{sec:num} confirm.

\subsection{Integrable Bounds}
\paragraph{Basic bound:}
We now consider integrable systems, which possess an additional set of quasi-local charges $Q_k$, where $Q_k$ has interaction length $k$. We assume that they are all translation invariant,
\beq
Q_k = \sum_x q_{x,(k)},
\eeq
where the operator $q_{x,(k)} = P^x q_{0,(k)} P^{-x} $ is supported on $k$ consecutive spins (being a sum of Pauli strings of the same length) and is traceless.
Note that
\begin{equation}
   \Tr(Q_kQ_l)=0\text{ if }k\neq l\,,
\end{equation}
since each term has the form $q_k q_l$, whose partial trace vanishes for $k\neq l$ because one of the two operators has a longer interaction length than the other. Hence the tensor products are orthogonal as well:
\begin{equation}
\label{eq:Q_norm}
    \bra (Q_k\otimes Q_k)(Q_\ell\otimes Q_\ell)\ket=\delta_{kl}\Tr(q^{2}_{0,(k)})^2L^2,
\end{equation}
Similarly,
\begin{equation}
    \bra \swap_A Q_k\otimes Q_\ell\ket=\frac{1}{d^2}\Tr_A(\Tr_B(Q_k)\Tr_B(Q_l))=\begin{cases}\frac{\Tr(q_{0,(k) }^2)}{d_A}(L_A-k+1)\text{ if }k=l\leq L_A
    \\0\text{ otherwise}
    \end{cases}
\end{equation}

Suppose the system is integrable in the conventional sense, with conserved charges $Q_k$ for all $k$ up to $k=L$. In deriving the Mazur bound, Then we must then sum over $k$ from $1$ to $L_A$, 
Since all these charges are mutually orthogonal, the bound is then given by
\begin{equation}
\label{eq:linear}
    \text{B}(\{Q_k\otimes Q_k\}_{k=1}^L)=\frac{1}{d_A^2}\sum_{k=1}^{L_A}\frac{(L_A-k+1)^2}{L^2}\approx \frac{1}{d_A^2} \frac{L_A^2}{L^2}\frac{L_A}{3}, \ L_A \gg 1.
\end{equation}

In deriving this formula, we assumed that each $Q_k$ is a sum of Pauli strings of the same length. The generalization to charges containing strings of several lengths is straightforward: it may change the numerical coefficients, but not the scaling.
Also, we assumed periodic boundary conditions. For open boundaries the normalization (\ref{eq:Q_norm}) of the charge has $(L-k+1)$. This also does not affect the scaling.

\paragraph{Improving the bound: products of charges}
The bound can be strengthened by considering products of different charges, provided we assume additional orthogonality properties.
Specifically, we assume that the $Q_n$ are ``sufficiently different'' that the trace of a product can be non-zero only when the charges are paired; in other words, the product of two charges cannot produce a third one, $Q_n Q_m \neq Q_k$. This is expected from the algebraic independence of the $Q_k$.
Effectively, this makes the traces ``Gaussian''. For example,
\beq
\langle Q_n Q_m Q_k Q_l \rangle = \delta_{nm}  \delta_{kl} \langle Q_n^2 \rangle \langle Q_k^2 \rangle
+ \text{[permutations]} .
\eeq
We can then consider a correction to the bound arising from taking products of different charges (we could of course also consider products of the same charge, but this is beyond the scope of this paper. For the Hamiltonian, this is done in the Supplementary Material, B.2).
We thus consider the set $\{\Sc^{n,p}\}$, defined by
\beq
\Sc^{n,p} = (Q^{n_1}_1 Q^{n_2}_2 \dots) \otimes (Q^{p_1}_1 Q^{p_2}_2
\dots) ,
\eeq
where each $n_i, p_i \in \{0,1\}$ indicates whether the corresponding charge is present.

We assume that
\beq
\bra \Sc^{n,p} \Sc^{\tilde{n},\tilde{p}} \ket
= \prod_i \delta_{n_i \tilde{n}_i} \delta_{p_i \tilde{p_i}} \bra Q_i^2 \ket^{n_i+p_i} .
\eeq
In fact, we need this property in a stronger form, after restricting each product to the subsystem $A$:
\beq
\bra \text{SWAP}_A \Sc^{n,p} \ket \equiv \frac{1}{d_A} \bra  \bra Q^{n_1}_1 Q^{n_2}_2 \dots \ket_B \bra Q^{p_1}_1 Q^{p_2}_2 \dots \ket_B \ket_A  = \frac{1}{d_A} \prod_i \delta_{n_i p_i} \bra \bra  Q_i \ket_B  \bra Q_i \ket_B \ket_A^{n_i},
\eeq
where $\bra \cdot \ket_B = \frac{1}{d_B} \Tr_B [\cdot]$.
Summing over all choices of $\{n_i\}$ then gives
\beq
\sum_{\{ n_i \}} \frac{\bra \bra Q_i  \ket_B \bra Q_i \ket_B \ket_A^{2 n_i}}{\bra Q_i^2 \ket^{2 n_i}} = \prod_i \l(1+ \frac{  \bra \bra  Q_i \ket_B \bra Q_i \ket_B \ket_A^{2}}{\bra Q_i^2 \ket^{2}} \r) .
\eeq
By the computation above, we get
\beq
\frac{  \bra \bra  Q_i \ket_B \bra Q_i \ket_B \ket_A^{2}}{\bra Q_i^2 \ket^{2}} = \frac{(L_A-i+1)^2}{L^2} , \quad 1 \le i \le L_A,
\eeq
and, as before, $\Tr(q_{0,(i)}^2)$ cancels. For $L_A \ll L$ we can approximate the product as 
\beq
\label{eq:exp}
\exp \l( \sum_{i=1}^{L_A} \log \l( 1 + \frac{(L_A-i+1)^2}{(L)^2} \r) \r) \approx \exp \l( \sum_{i=1}^{L_A} \frac{(L_A-i+1)^2}{(L)^2} \r)  \approx \exp \l( \frac{1}{3} \frac{L_A^3}{L^2}  \r)\,,
\eeq
where in the last approximation we assumed additionally $L_A \gg 1$.

\subsection{Formal statement}
\label{sec:formal}
In this subsection, we use the examples above to formulate a general statement, in order to emphasize which property of the conserved charges is actually needed.
Motivated by the form of our bound and by the explicit conserved charges of known integrable models, we adopt the following definition of a quasi-local conserved charge $Q_k$:

\begin{definition}[Quasi-local conserved charge $Q_k$]
An operator $Q_k$ is called a quasi-local conserved charge of length $k$ and weight $c_k$ if $Q_k$ commutes with the Hamiltonian, and there exists a positive constant $c_k$, independent of the system size, such that for every contiguous subsystem $A$ with $k\le L_A$ the following ratio is bounded from below:
\beq
\label{eq:def}
\frac{ \Tr_A [\Tr_{\bar{A}} (Q_k) \Tr_{\bar{A}} (Q_k) ]}{\Tr Q_k^2} \ge c_k d_B \frac{L_A-k+1}{L}
\eeq
\end{definition}

This notion of quasi-locality is non-trivial: an all-to-all operator effectively has infinite length, since its partial trace over a typical subsystem vanishes. The definition allows $Q_k$ to contain Pauli strings of different lengths.
Other notions of quasi-locality exist in the literature. For example, refs. \cite{Mierzejewski:2015ctc, Pereira_2014} found a remarkable family of charges $I_k$ in the XXX and XXZ Heisenberg model such that, in the limit $L \ra \infty$, $\Tr I_k^2 \sim L$, while $\bra P I_k \ket = O(L^0)$ for any Pauli string $P$. It would be very interesting to understand their contribution to the Mazur bound on the operator purity. Presumably it takes the form of eq.~(\ref{eq:def}), possibly with a different power of $(L_A-k+1)/L$.

Applying this definition to the Mazur bound (\ref{eq:intro_mazur}) immediately yields
\begin{proposition}[Lower bound on operator purity]
Suppose that a one-dimensional Hamiltonian system of $L$ qubits has a family of Hilbert--Schmidt orthogonal quasi-local charges $Q_k$ with lengths up to $M$. Then, for any contiguous subsystem $A$ of length $L_A \le M$, the infinite-time average of the operator purity $\Tr \rho_{A_1 A_2}^2$ satisfies
\beq
d_A^2 \times \overline{\Tr \rho_{A_1 A_2}^2} \ge (\min c_k^2 )\sum_{k=1}^{L_A} \frac{(L_A-k+1)^2}{L^2} .
\eeq
If the $c_k$ are bounded from below, the right-hand side grows linearly in $L_A$ for large $L$ and $L_A$ at fixed ratio $L_A/L$.
\end{proposition}

\section{Numerics}
\label{sec:num}
We test our predictions with exact diagonalization in three models: the XXX and XYZ spin chains and the mixed-field Ising model. To show that our bounds apply more generally, we impose open boundary conditions.

For the XYZ chain,
\beq
H_{XYZ} = \sum_i J_x X_i X_{i+1} + J_y Y_{i} Y_{i+1} + J_z Z_i Z_{i+1},
\eeq
we break integrability with a next-nearest-neighbor term
\beq
H_{NN} = \sum_i B_x X_i X_{i+2} + B_y Y_{i} Y_{i+2} + B_z Z_i Z_{i+2} .
\eeq
For the XYZ chain we take $(J_x,J_y,J_z) = (1,2,3)$ and $(B_x, B_y, B_z)= (1,1,1)$, and for the XXX chain $(J_x,J_y,J_z) = (1,1,1)$ and $(B_x,B_y,B_z)=(2,2,2)$.

For the Ising model, integrability is broken by adding a non-zero $h_z$ magnetic field:
\beq
H_{Ising} = \sum_i \l( -Z_i Z_{i+1} + h_z Z_i + h_x X_i \r) .
\eeq
We take $h_x=1$ and set $h_z$ to either $0$ or $1/2$.

\begin{figure}[h!]
\begin{minipage}{0.5\textwidth}
    \centering
    \includegraphics[scale=0.8]{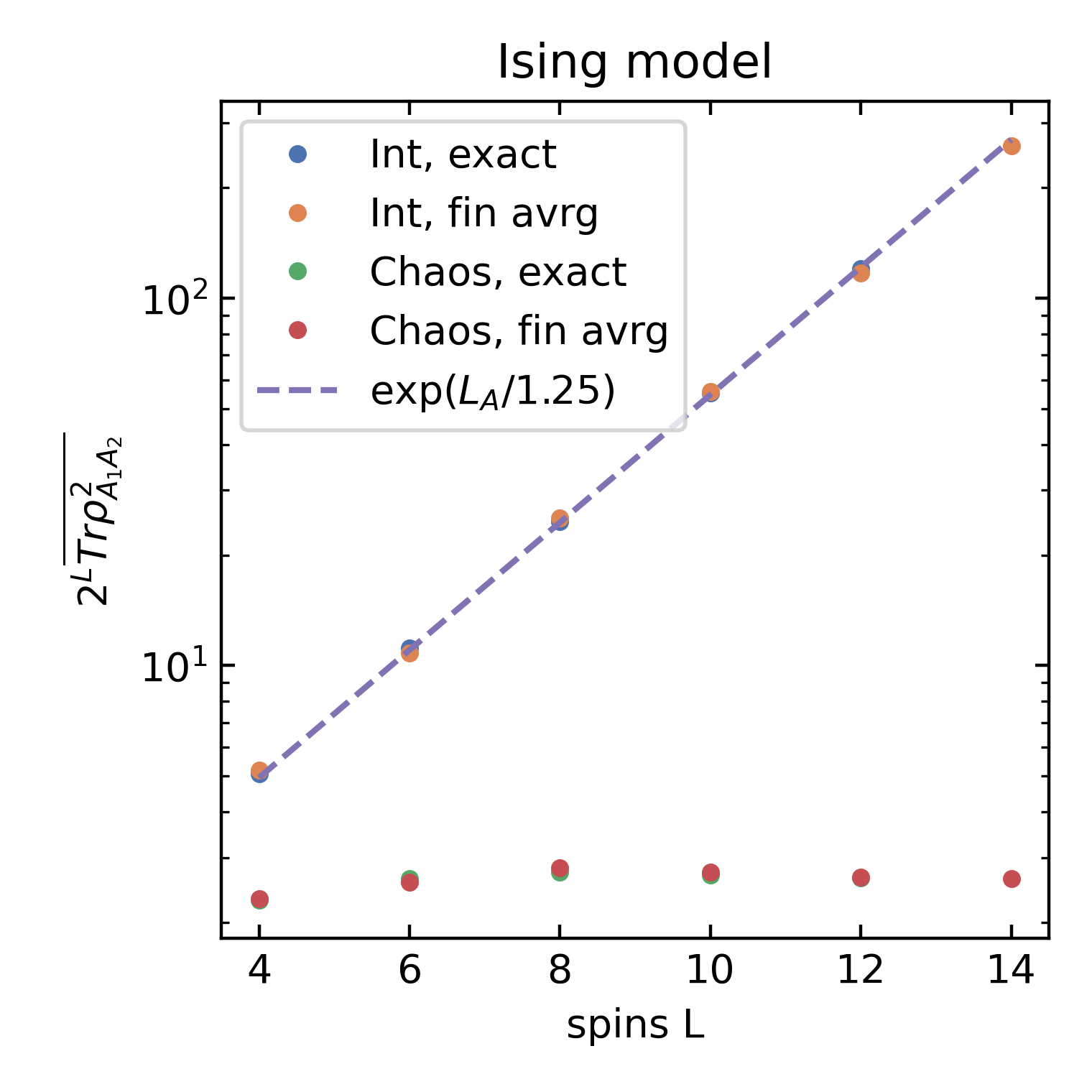}
\end{minipage}
\begin{minipage}{0.5\textwidth}
    \centering
     \includegraphics[scale=0.8]{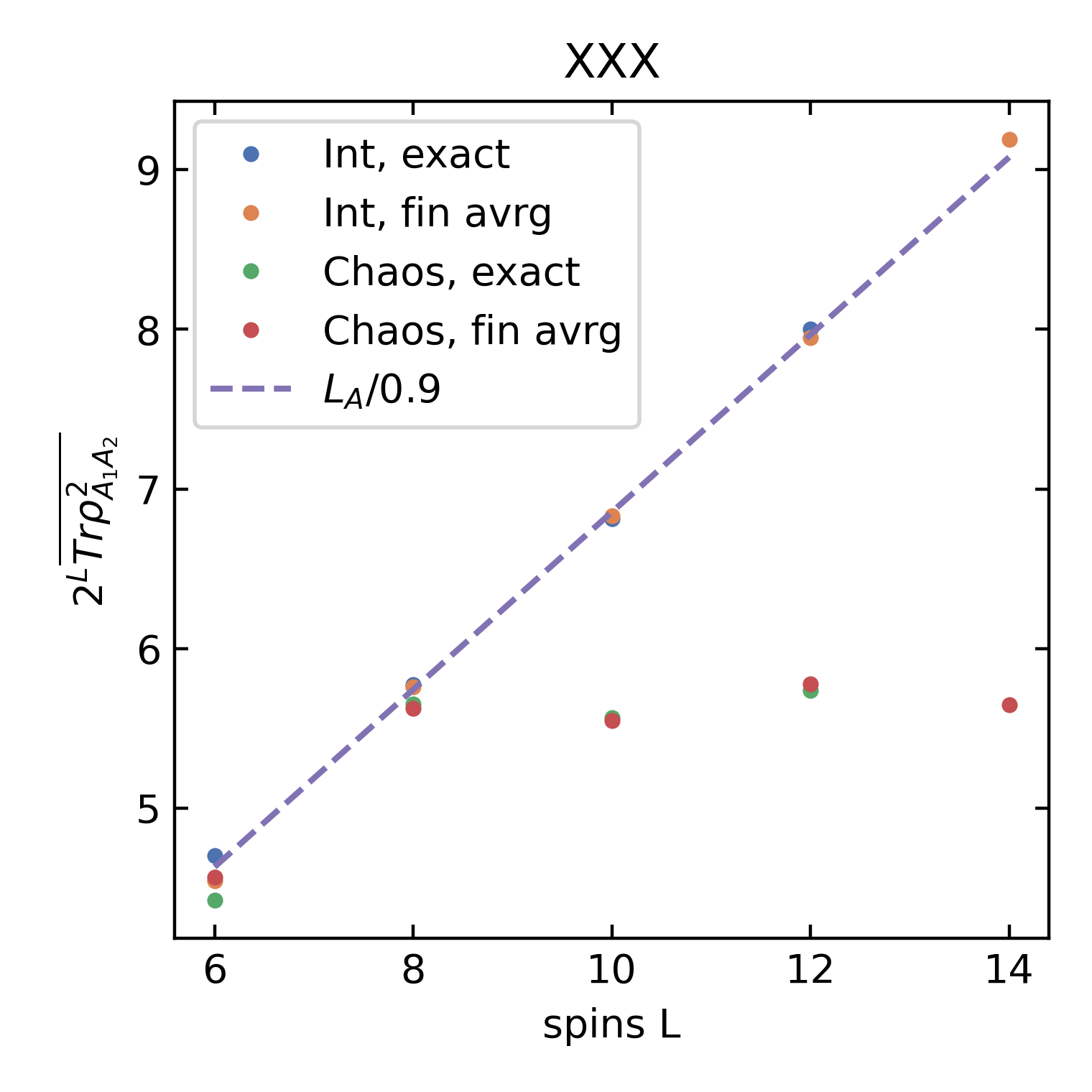}
\end{minipage}

\begin{center}
    \centering
     \includegraphics[scale=0.8]{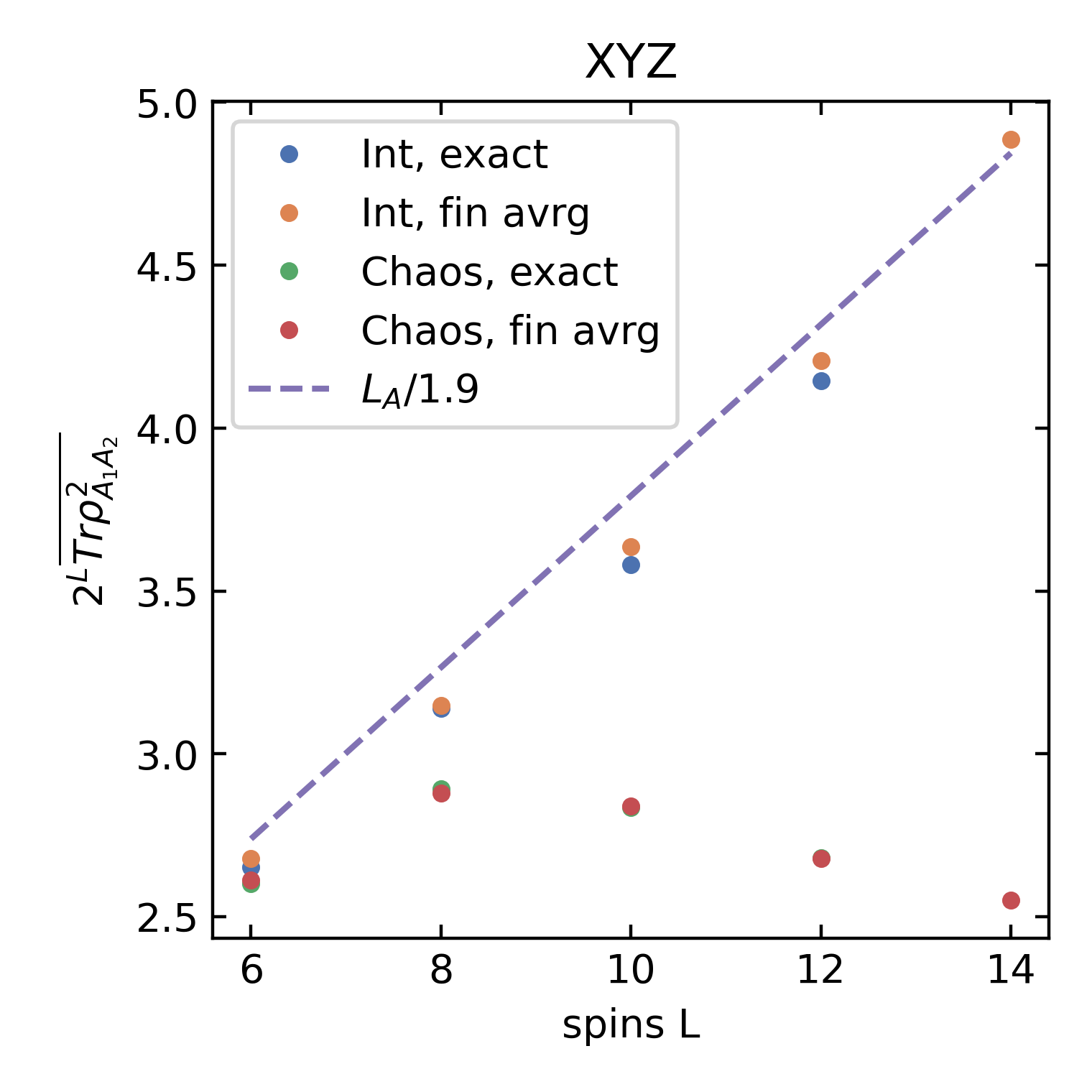}
\end{center}
\caption{Numerical results for the three models described in the main text, with $L_A=L/2$ throughout. Dots show the numerical data and dashed lines are fits. Error bars from the finite-sample average are smaller than the dots.}
\label{fig:num}
\end{figure}

We compute the time average (\ref{eq:time_avg}) in two ways.
First, we can integrate over time analytically. This produces overlaps of the SWAPs with the energy eigenstates $|E_n \ket, |E_m \ket, |E_k\ket, |E_l\ket$, which we then sum over $n,m,k,l$ subject to the condition\footnote{Numerically, such an exact condition cannot be imposed on floating-point numbers. In practice we implement it with a tolerance of $10^{-8}$, which is equivalent to exact time integration up to times of order $10^8$.} $E_n - E_m = E_k - E_l$. This procedure is expensive, but feasible up to $L=12$ spins.
Second, we can sample a finite number of time points at late times and compute the empirical average; specifically, we sample times between $L^2$ and $4 L^2$. This method is significantly faster and easily reaches $L=14$. The agreement between the two methods supports our claim that $\Tr \rho^{2}_{A_1 A_2}$ saturates at times polynomial in the system size.
An important caveat is the error due to finite sampling. We assume that the samples are independent and average them to reduce the error, which is justified by the very good agreement between the two methods. It would be interesting to understand the statistics of the late-time fluctuations in more detail. In agreement with previous observations in the literature \cite{Fortes:2019frf}, we find empirically that the fluctuations are significantly larger in the integrable case than in the chaotic case.

Our results are presented in Figure \ref{fig:num}. They agree with our expectations for all the models considered: in the chaotic case the average is constant, while in the integrable case it grows at least linearly. In fact, it grows considerably faster, and we comment on this now.
For the Ising model we observe a very rapid exponential growth, with a coefficient in the exponent larger than in eq.~(\ref{eq:exp}). We attribute this to the fact that multiple copies of the Ising model have an \textit{extensive} number of additional conserved charges \cite{wip}, similar to free fermion case \cite{Lastres:2026nbx,Vardhan:2024wxb}.
For the XYZ and XXX chains, we observe a linear growth with coefficients larger than predicted by eq.~(\ref{eq:linear}), and we do not see the exponential growth obtained under rather strong assumptions in eq.~(\ref{eq:exp}).
We nevertheless expect this growth to be explained by eq.~(\ref{eq:exp}): our numerics use $L_A/L=1/2$, for which the actual asymptotics of the discrete sum is\footnote{For that one needs to use the open boundary condition formula which is $\sum_{k=1}^{L_A} (L_A-k+1)^2/(L-k+1)^2$} $0.11 L_A$, and they reach at most $L_A=7$, where the exponential is almost indistinguishable from a straight line.
In this regime, the prefactor of the exponential also affects the linear coefficient, and since $L_A=L/2$, the contribution of the SWAP operator becomes important. Finally, as mentioned above, refs. \cite{Mierzejewski:2015ctc, Pereira_2014} uncovered additional non-local conserved charges in the XXX/XXZ model which contribute to the generalized Gibbs ensemble \cite{Ilievski:2015jhc}, and we expect them to contribute to our bound as well.

\section{Discussion}
In this paper, we systematically studied the late-time residual value of the operator purity $\Tr \rho_{A_1 A_2}^2$ (the exponentiated second R\'enyi entropy).
Using the Mazur bound (\ref{eq:intro_mazur}) and the Suzuki equality, we argued that this value behaves very differently in integrable and chaotic systems: in chaotic systems $d_A^2 \Tr \rho_{A_1 A_2}^2$ saturates to a value that depends only on the ratio $L_A/L$, eq.~(\ref{eq:chaos_val}), whereas in integrable systems it grows at least linearly in $L_A$ at fixed $L_A/L$, eq.~(\ref{eq:int_val}).
We tested these predictions numerically in several one-dimensional systems and found good agreement; in the integrable case the growth even exceeds our naive bounds.

We outline several directions for future research, including a key conjecture about the saturation time:
\begin{itemize}
\item As mentioned above, we expect the operator entropy to saturate at times polynomial in the system size. This matters both conceptually, for understanding the time scales on which chaos and integrability can be distinguished, and practically, for the numerical computation described in Section \ref{sec:num}. So far we have provided only preliminary numerical evidence for this. We hope to address the question more rigorously using the eigenstate thermalization hypothesis (ETH) \cite{Srednicki:1994mfb,PhysRevA.43.2046}(including subsystem ETH \cite{Dymarsky:2016ntg} and the corrections relevant for OTOCs \cite{Foini:2018sdb,Brenes:2021bjr}) together with more precise numerical checks.
Our sampling procedure in Section \ref{sec:num} also \textit{assumed} that the late-time fluctuations are independent and random; we hope to investigate these fluctuations in detail in the future.

\item To apply the Mazur bound, we studied the operator purity for the same region $A$ before and after the evolution, since it can then be written as an autocorrelation function of a SWAP operator. Operator purity can be defined for any pair of regions, and different configurations sometimes carry interesting physical information \cite{Asplund:2015eha,Nie:2018dfe}. Although this quantity remains positive, it is not obvious how to write a Mazur bound for it. Also, it would also be interesting to see whether higher R\'enyi entropies lead to new physical insights.

\item Ref.~\cite{Fortes:2019frf} observed empirically that late-time OTOCs fluctuate more strongly in integrable systems than in chaotic ones. We believe that this observation can be explained with our methods by writing a Mazur-type bound for the variance of the OTOC. Such a bound again involves a sum over the conserved charges and is therefore naturally larger for integrable systems; this approach should also predict different scalings with the system size.

\item Rather than the sum over OTOCs that defines the operator entropy, it would be interesting to study the OTOCs of individual operators, which can carry additional physical information. For example, refs.~\cite{Garcia:2024scq,Jackson:2026gmd} showed that OTOCs can serve as a measure of non-stabilizerness (the so-called ``magic'').

\item In this paper, we studied operator entropies of the evolution operator $e^{-i H t}$, which can be written as sums of infinite-temperature OTOCs. Physically, it would be instructive to study the finite-temperature case. Ref. \cite{Anand:2021xet} has a proposal for a finite-temperature generalization and it would be interesting to see if it admits an analytic lower bound. 
 
\end{itemize}

\section*{Acknowledgment}
We are indebted to Shreya~Vardhan for collaboration at the early stages and for numerous illuminating discussions. We would like to thank Anatoly~Dymarsky, Alexei~Kitaev, Shota~Komatsu,  John~Preskill for comments.

\appendix 

	\section{Mazur bound}
	\label{app:mazur}
	Let us recall the proof of the Mazur bound. Consider a hermitian $Y$ and insert a resolution of the identity to write
	\begin{equation}
		\bra Y(t) Y(0)\ket=\sum_{n,m}\rho_n\bra n|Y(t)|m\ket\bra m|Y(0)|n\ket\,,
	\end{equation}
	where we assumed for simplicity that $\rho$ is diagonal in the energy eigen basis. Then
	\begin{equation}
		\bra n|Y(t) |m\ket=e^{i(E_n-E_m)t} Y_{nm}\,,
	\end{equation}
	where $Y_{nm}=\bra n|Y(0)| m\ket$: as $t\ra\infty$ the computation thus picks out only the elements of $Y_{nm}$ with $E_n=E_m$. Then we get the bound
	\begin{equation}
		\overline{\bra Y(t)Y(0)\ket}=\sum_{E_n=E_m}\rho_n Y_{nm} Y_{mn}\geq 0\,,
	\end{equation}
	since this matrix is manifestly positive semi-definite (being the square of another matrix, multiplied by the positive definite matrix $\rho$). 
	Now suppose the system has some Hermitian conserved charges $\{Q_k\}_{k=1}^{N}$. Then one can apply the above inequality to $\tilde{Y} (t)= Y(t) - \sum_k z_k Q_k$ where $z_k$ are arbitrary real numbers. The bound then takes the form
	\begin{equation}
		\overline{\bra \tilde{Y}(t)\tilde{Y}(0)\ket}=\overline{\bra Y(t)Y(0)\ket}+\sum_{ij}z_iz_j \bra Q_i Q_j\ket-2\sum_{i}z_i\bra Y Q_i\ket\geq0\,,
	\end{equation}
	where operators without explicit time-dependence are defined at $t=0$, and we used that $Q_i$ is preserved and that the state is time translation-invariant. Extremizing this bound wrt. $z_i$ sets $z_i= M_{ij}\bra Y Q_j\ket$ where $M_{ij}=\bra Q_i Q_j \ket^{-1}$ means the inverse matrix of $\bra Q_i Q_j \ket$ (whose entries are labeled by $i$ and $j$), and we thus obtain eq. (\ref{eq:thebound}).

	\section{Bounds from Powers of Translation and the Hamiltonian}
	In this appendix, we treat the bound that follows from the space- and time-invariance of the system \eqref{eq:Hamil}. We will consider independently the bound that follows from taking powers of the translation operator $P$ as well as the Hamiltonian $H$, the latter providing corrections to the bound \eqref{eq:BHH} derived in the main text. As we shall see, they provide corrections that are down by factors of $r_A=L_A/L$ or (even more suppressed, factors of $L^\alpha/d^\beta_C$ for some subsystem $C$).
    In a more complete analysis, we should of course combine these together, and also consider charges of the form $P^n\otimes H^m$. However, we also check that at least for $n=m=1$, such corrections are parametrically small in $a=\sqrt{L}/d$.
	\subsection{Bound from powers of translation}
	\label{app:trans}
	Since the Hamiltonian is translation invariant, we can use the conservation of powers of $P$. 
	We will assume that each site of the system consists of $N$-bit system, such that $d=N^L$, and work at infinite temperature. 
    
    Since $P^n$ is not Hermitian, we need to use the appropriate definition of the matrix $M$, namely $M_{nm,kl}=\left<(P^n\otimes P^m)^{\dagger}\,(P^k\otimes P^l)\right>^{-1}$, where $(P^n)^\dagger=P^{-n}$. We will produce an upper bound on $B$ by using
    \begin{equation}
    \begin{split}
        B\left(\{P^n\otimes P^m\}_{n,m=0}^{L-1}\right)&=\sum_{n,m,k,l=0}^{L-1}\left<\SWAP_A\,P^n\otimes P^m\right>^* M_{nm,kl}\left<\SWAP_A\,P^k\otimes P^l\right>
        \\
        &\leq \sum_{n,m=0}^{L-1}|\left<\SWAP_A\,P^n\otimes P^m\right>|^2\lambda_{\text{max}}\,,
        \end{split}\label{eq:PBound}
    \end{equation}
    where $\lambda_{\text{max}}$ is the largest eigenvalue of $M$. While producing an upper bound on $B$ might seem counterproductive (after all, it is itself a \emph{lower bound} on the two-point function), this allows us to bound the corrections to \eqref{idbound}, which must be parametrically small.
    
    We will use the following identity extensively
    \begin{equation}
        \Tr(P^n)=N^{\gcd(L,n)}\,,
    \end{equation}
	where $\gcd(x,y)$ denotes the greatest common divisor between $x$ and $y$. Since $(P^n)^{\dagger}=P^{-n}$, this implies that the diagonal entries of the Gram matrix are $1$. Since $\gcd(L,n-k)\leq L/2$ for $n\neq k$, it also allows us to bound the non-diagonal entries. Using Gershgorin theorem, this leads to the following upper bound on the maximal eigenvalue of $M$
	\begin{equation}
		\lambda_{\text{max}} \leq \left[1-\frac{L-1}{\sqrt{d}}\right]^{-2}\,.\label{eq:MTrans}
	\end{equation}
    This upper bound is not very tight, but suffices for our purposes. 
    
    Similarly, we can bound the overlaps. Here, we will see that the terms with $n=m=0$ contribute very differently than the rest, and it is instructive to bound the operator where these contributions have been subtracted
	\begin{equation}
		\tilde{\SWAP}_A=\SWAP_A-\frac{1}{d_A}\mathbb{I}\otimes \mathbb{I}\,.\label{eq:tdswap}
	\end{equation}
    The overlaps with this operator take the form
    \begin{equation}
        \left<\tilde{\SWAP}_A\,P^n\otimes P^m\right>=\frac{1}{d^2}\left[\Tr_A(\Tr_B(P^n)\Tr_B(P^m))-\frac{\Tr(P^n)\Tr(P^m)}{d_A}\right]=\frac{1}{d^2}\Tr_A(\mathbb{P}_n\mathbb{P}_m)\,,
    \end{equation}
    where we defined
    \begin{equation}
        \mathbb{P}_n\equiv \Tr_B(P^n)-\frac{\Tr(P^n)}{d_A}\mathbb{I}_A=\Tr_B(P^n)-\frac{N^{\gcd(n,L)}}{d_A}\mathbb{I}_A\,.
    \end{equation}
    Since the trace is an inner product, to bound $|\Tr_A(\mathbb{P}_n\mathbb{P}_m)|^2$ appearing \eqref{eq:PBound}, it suffices to bound $\Tr_A(\mathbb{P}^\dagger_n\mathbb{P}_n)$ and $\Tr_A(\mathbb{P}^\dagger_m\mathbb{P}_m)$. The $\mathbb{P}_n$ have the property that they vanish whenever $\gcd(n,L)\geq L_A$, while the remaining ones satisfy
    \begin{equation}
        \Tr_A(\mathbb{P}_n^\dagger\mathbb{P}_n)=d_A-\frac{N^{2\gcd(L,n)}}{d_A}\leq d_A\label{eq:Pnbound}
    \end{equation}
    Combining \eqref{eq:MTrans} and \eqref{eq:Pnbound} and adding back in the contribution coming from $\mathbb{I}\otimes \mathbb{I}$ then allows us to establish
    \begin{equation}
    \begin{split}
        B\left(\{P^n\otimes P^m\}_{n,m=0}^{L-1}\right)&\leq\frac{1}{d_A^2}+\sum_{n,m=1}^{L-1}\left(\frac{d_A}{d^2}\right)^2{\lambda_{\text{max}}}
        \\
        &\leq \frac{1}{d_A^2}\left[1+\frac{(L-1)^2}{d_B^4\left(1-\frac{L-1}{\sqrt{d}}\right)^2}\right]=\frac{1}{d_A^2}[1+O(L^2/d_B^4)]
    \end{split}
    \end{equation}
    and the corrections to the contribution from the identity  \eqref{idbound} are thus exponentially suppressed.
	\subsection{Bounds from powers of the Hamiltonian}
	\label{app:Hpower}
	We now consider operators of the form $H^n\otimes H^m$, with powers up to $n_{max}$. As we shall see, the lower bound on $d_A^2 \Tr \rho^2_{A_1 A_2}$ is only a function of the ratio $r_A=L_A/L$, so it will not grow with $L$ at fixed ratio.
	
	As a warm-up, let us first consider the case $m=0$, i.e. operators of the form $H^n\otimes \mathbb{I}$. We can use the fact that for any operator $Y$ acting on a single copy of the system, we have the identity 
	\begin{equation}
		\left<\SWAP_A\, Y\otimes\mathbb{I}\right>=\frac{1}{d^2}\Tr_A(\Tr_B(Y)\Tr_B(\mathbb{I}))=\frac{d_B}{d^2}\Tr(Y)=\frac{1}{d_A}\left<\mathbb{I}\otimes\mathbb{I}\, Y\otimes\mathbb{I}\right>\,.
	\end{equation}
	This implies that $\tilde{\SWAP}_A$ as defined in \eqref{eq:tdswap} is orthogonal to all operators of the form $Y\otimes \mathbb{I}$. Thus, the only contributions to the bound arising from $H^n\otimes\mathbb{I}$ come from the term with $n=0$, and we obtain
	\begin{equation}
		B\left(\left\{H^n\otimes\mathbb{I}\right\}_{n=0}^{n_{max}}\right)=B\left(\left\{\mathbb{I}\otimes H^n\right\}_{n=0}^{n_{max}}\right)=B(\mathbb{I})=\frac{1}{d_A^2}
	\end{equation}
	Next, we can now tackle the cases with $n,m\neq 0$. We will consider the thermodynamic limit $\ell/L\rightarrow 0$, where $\ell$ is the interaction length of the Hamiltonian. Define 
	\begin{equation}
	\mathbb{H}=H\sqrt{\frac{N^\ell}{\Tr(h_0^2)L}}\,,
	\end{equation}
	which is normalised such that $\left<\mathbb{H}\otimes\mathbb{I}\,\mathbb{H}\otimes\mathbb{I}\right>=1$ in the limit. 
    
    The traces of higher powers of $\mathbb{H}$ reduce to Gaussian moments. For example 	\begin{equation}
		\frac{1}{d} \bra H^4 \ket= L^4 \bra h_0 \ket^4+3L^2 \bra h_0^2 \ket^2+4L^2\bra h_0^3 \ket \bra h_0 \ket+L \bra h_0^4 \ket+O(L)\underset{L\ra\infty}{\sim}3L^2 \bra h_0^2 \ket^2\,,
	\end{equation}
	where we used that $\Tr(h_0)=0$ and the coefficients follow from the number of ways we can place the individual $h_0$ such that they overlap. This leads to the expression
		\begin{equation}
		\frac{1}{d}\Tr\left(\mathbb{H}^{2n}\right)=(2n-1)!!\,,
	\end{equation}
    which are the moments of the Gaussian distribution.
	Gaussian moments are orthogonalised by the normalised Hermite polynomials, such that
	\begin{equation}
		\mathbb{H}_n\equiv\frac{\text{He}_n(\mathbb{H})}{\sqrt{n!}}\Rightarrow \left<\mathbb{H}_n\otimes\mathbb{H}_m\,\mathbb{H}_k\otimes\mathbb{H}_l\right>=\delta_{nk}\delta_{ml}\,.\label{eq:Hnortho}
	\end{equation}
    
	That the higher powers are Gaussian makes it possible to compute the overlap with the swap operator. We can decompose $\mathbb{H}$ into $\mathbb{H}_A+\mathbb{H}_B+\mathbb{H}_{\text{mix}}$, where $\mathbb{H}_A$ and $\mathbb{H}_B$ act only on subsystem $A$ and $B$, respectively, and $\mathbb{H}_{\text{mix}}$ includes the interaction between the two subsystems. In the thermodynamic limit, the contribution from $\mathbb{H}_{\text{mix}}$ can be safely ignored. $\exp(u\mathbb{H})$ is clearly a generating function for $\mathbb{H}^n$, and the Gaussian behaviour of the moments of $\mathbb{H}$ then allow us to write
	\begin{equation}
		\begin{split}
		&\Tr\left(\SWAP_A e^{u\mathbb{H}}\otimes e^{v\mathbb{H}}\right)=\Tr_A(\Tr_B(e^{u\mathbb{H}})\Tr_B(e^{v\mathbb{H}}))
		\\
		=&\Tr_A(e^{(u+v)\mathbb{H}_A})\Tr_B(e^{u\mathbb{H}_B})\Tr_B(e^{v\mathbb{H}_B})=d_Ad_B^2\exp\left(\frac{(u+v)^2}{2}r_A\right)\exp\left(\frac{u^2}{2}r_B\right)\exp\left(\frac{v^2}{2}r_B\right)\,,
		\end{split}
	\end{equation}
	where we replaced the exponentials of $\mathbb{H}$ with a corresponding Gaussian. Given \eqref{eq:Hnortho}, it is helpful to change basis from $\mathbb{H}^n$ to $\mathbb{H}_n$, which is straightforward using that $e^{u\mathbb{H}-u^2/2}$ is a generating function for $\mathbb{H}_n$. A similar computation then yields
	\begin{equation}
		\Tr\left(\SWAP_A e^{u\mathbb{H}-u^2/2}\otimes e^{v\mathbb{H}-v^2/2}\right)=d_Ad_B^2e^{uvr_A}\,.
	\end{equation}
	From this, one obtains the bound
	\begin{equation}
		B(\left\{H^n\otimes H^m\right\}_{n,m=0}^{n_{max}})=B(\left\{\mathbb{H}_n\otimes\mathbb{H}_m\right\}_{n,m=0}^{n_{max}})=\frac{1}{d_A^2}\sum_{n=0}^{n_{max}}r_A^{2n}\cong\frac{1}{d_A^2}\frac{1}{1-r_A^2}\,.
	\end{equation}
	where we used that fact that in the thermodynamic limit $n_{max} \ra \infty$. This bound only depends on $r_A$ as stated in the main text. 
	\subsection{Combining the Hamiltonian and Translation}
	We have also checked what we get if we consider combining $\mathbb{H}$ and $P$. Here we will consider them only to the first power, but we expect the result to hold more generally. In the basis $P\otimes P,P\otimes \mathbb{H},\mathbb{H}\otimes P,\mathbb{H}\otimes \mathbb{H}$, we find that $M^{-1}$ takes the form
	\begin{equation}
		M^{-1}\sim\begin{pmatrix}
			1&a&a&a^2\\
			a&1&a^2&a\\
			a&a^2&1&a
			\\
			a^2&a&a&1
		\end{pmatrix}
	\end{equation}
	where $a=\sqrt{L}/d<\!\!\!<1$. Thus, the off-diagonal elements are highly suppressed, and any correction to the bound coming from mixed operators is down by factors of $a$.
\printbibliography

\end{document}